\documentclass[aps,prd,twocolumn,amsmath,amssymb,amsfonts,nofootinbib,superscriptaddress]{revtex4-1}
\usepackage[utf8]{inputenc}
\usepackage{graphicx}
\usepackage{subfigure}
\usepackage{datatool}
\usepackage[scientific-notation=true]{siunitx}
\usepackage[dvipsnames]{xcolor}
\usepackage{lineno}
\usepackage{xspace}
\usepackage{orcidlink}
\usepackage{url}

\usepackage{breakurl}
\usepackage{lineno}
\usepackage{longtable}
\usepackage[shortlabels]{enumitem}
\usepackage[capitalise]{cleveref}

\def\gwh{GW\xspace}

\def\gws{GWs\xspace}

\def\dm{DM\xspace}
\def\dmh{DM\xspace}
\def\pbh{PBH\xspace}
\def\pbhs{PBHs\xspace}

\def\fh{frequency-Hough\xspace}
\def\lvk{LIGO, Virgo and KAGRA\xspace}

\def\invkpccubedyr{kpc$^{-3}$yr$^{-1}$\xspace}
\def\pn{PN\xspace}

\def\GFH{\emph{Generalized frequency-Hough}\xspace}

\newcommand{\bea}{\begin{eqnarray}}
\newcommand{\eea}{\end{eqnarray}}
\newcommand{\be}{\begin{equation}}
\newcommand{\ee}{\end{equation}}

\newcommand{\avgVT}{\ensuremath{\left\langle VT \right\rangle}}

\newcommand{\TFFT}{T_\text{FFT}}

\newcommand{\Tobs}{T_\text{PM}}
\newcommand{\fpbh}{f_\text{PBH}}

\newcommand{\fsup}{f_\text{sup}}
\newcommand{\Mc}{\mathcal{M}}

\newcommand{\ssm}{SSM\xspace}
\newcommand{\msun}{\ensuremath{M_\odot}\xspace}

\newcommand{\mpbh}{m_\text{PBH}}

\def\erfc{\mathrm{erfc}}

\newcommand{\DaysDataAnalyzed}{126 days\xspace}

\graphicspath{{./figures/}}

\begin{document}

\title{Gravitational-wave constraints on planetary-mass primordial black holes \\ using LIGO O3a data}

\author{Andrew L. Miller\,\orcidlink{0000-0002-4890-7627}}
\email{andrew.miller@nikhef.nl}
\affiliation{Nikhef -- National Institute for Subatomic Physics,
Science Park 105, 1098 XG Amsterdam, The Netherlands}
\affiliation{Institute for Gravitational and Subatomic Physics (GRASP),
Utrecht University, Princetonplein 1, 3584 CC Utrecht, The Netherlands}
\author{Nancy Aggarwal}
\email{nqaggarwal@ucdavis.edu}
\affiliation{University of California, Davis, Department of Physics,
Davis, CA 95616, USA}
\author{Sébastien Clesse}
\affiliation{Service de Physique Th\'eorique, Universit\'e Libre de Bruxelles, Boulevard du Triomphe CP225, B-1050 Brussels, Belgium}
\author{Federico De Lillo}
\affiliation{Université catholique de Louvain, B-1348 Louvain-la-Neuve, Belgium}
\author{Surabhi Sachdev}
\affiliation{School of Physics, Georgia Institute of Technology, Atlanta, GA 30332, USA}
\author{Pia Astone}
\affiliation{INFN, Sezione di Roma, I-00185 Roma, Italy}
\author{Cristiano Palomba}
\affiliation{INFN, Sezione di Roma, I-00185 Roma, Italy}
\author{Ornella J. Piccinni}
\affiliation{OzGrav, Australian National University, Canberra, Australian Capital Territory 0200, Australia}
\author{Lorenzo Pierini}
\affiliation{INFN, Sezione di Roma, I-00185 Roma, Italy}

\date{\today}

\begin{abstract}

Gravitational waves from sub-solar mass inspiraling compact objects would provide almost smoking-gun evidence for primordial black holes (PBHs). We perform the first search for inspiraling planetary-mass compact objects in equal-mass and highly asymmetric mass-ratio binaries using data from the first half of the LIGO-Virgo-KAGRA third observing run. Though we do not find any significant candidates, we determine the maximum luminosity distance reachable with our search to be of $\mathcal{O}(0.1-100)$ kpc, and corresponding model-independent upper limits on the merger rate densities to be $\mathcal{O}(10^{3}-10^{-7})$ kpc$^{-3}$yr$^{-1}$ for systems with chirp masses of $\mathcal{O}(10^{-4}-10^{-2})M_\odot$, respectively. 
Furthermore, we interpret these rate densities as arising from PBH binaries and constrain the fraction of dark matter that such objects could comprise. {For equal-mass PBH binaries, we find that these objects would compose less than 4-100\% of DM for PBH masses of $10^{-2}M_\odot$ to $2\times 10^{-3}M_\odot$, respectively. For asymmetric binaries, assuming one black hole mass corresponds to a peak in the mass function at 2.5$M_\odot$, a PBH dark-matter fraction of 10\% and a second, much lighter PBH, we constrain the mass function of the second PBH to be less than 1 for masses between $1.5\times 10^{-5}M_\odot$ and $2\times 10^{-4}M_\odot$.}
{Our constraints, released on Zenodo \cite{miller_2024_10724845}, are robust enough to be applied to \emph{any} PBH or exotic compact object binary formation models, and complement existence microlensing results}. More details about our search can be found in our companion paper  \cite{Miller:2024jpo}.

\end{abstract}

\maketitle

\section{Introduction}

The detection of low-spinning black holes by \lvk ~\cite{aasi2015advanced,acernese2014advanced,Abbott:2016blz,TheLIGOScientific:2016pea,Abbott:2016nmj,Abbott:2017vtc,Abbott:2017oio,Abbott:2017gyy,LIGOScientific:2018mvr,LIGOScientific:2020stg,Abbott:2020uma,Abbott:2020khf,Abbott:2020tfl,Abbott:2020mjq} has renewed interest in primordial black holes (\pbhs) as dark-matter (\dmh) candidates ~\cite{Bird:2016dcv,Clesse:2016vqa,Sasaki:2016jop}. Depending on when and how \pbhs formed in the early universe~\cite{Carr:2019kxo,Byrnes:2018clq,Jedamzik:2020ypm,Jedamzik:2020omx,DeLuca:2020agl,Escriva:2022duf}, they could have any mass between $\sim [10^{-18},10^9]\msun$, and could comprise a fraction $\fpbh$ ~\cite{Sasaki:2016jop,Ali-Haimoud:2017rtz,Hall:2020daa,DeLuca:2020jug} or all of \dm ~\cite{Bird:2016dcv,Clesse:2016vqa,clesse2018seven,Carr:2019kxo,Jedamzik:2020ypm,Jedamzik:2020omx,Boehm:2020jwd,DeLuca:2020agl}. Such a wide mass range necessitates different probes of \pbhs, one of which is through gravitational-wave (GW) emission.
However, there is ambiguity between astrophysical and primordial formation mechanisms to explain observations of black holes above a solar mass \cite{Clesse:2016vqa,Clesse:2020ghq,Takhistov:2020vxs}; hence, it is worthwhile to search for \gws from inspiraling compact objects below a solar mass, whose origins almost certainly would be primordial \cite{LISACosmologyWorkingGroup:2023njw,Yamamoto:2023tsr}. 

Despite a generic feature in the \pbh mass function that appears at $10^{-5}M_\odot$~\cite{Niemeyer:1997mt,Jedamzik:1996mr,Byrnes:2018clq,Carr:2019kxo}, 
much interest so far has focused on sub-solar mass (\ssm) \pbhs with masses between $[0.1,1]\msun$ using matched filtering, the ideal, computationally-intensive signal processing technique that correlates a huge number of waveforms with the data  \cite{LIGOScientific:2018glc,abbott2019search,Nitz:2020bdb,horowitz2020search,Nitz:2021vqh,LIGOScientific:2021job,LIGOScientific:2022hai}, with one exception 
\cite{Ebersold:2020zah}. 
While these searches have yielded no significant \gwh events\footnote{One search, however, claims to have seen a \ssm low-significance candidate \cite{Phukon:2021cus,Prunier:2023cyv}.}, they have placed stringent upper limits on $\fpbh\lesssim\text{few \%}$. 

However, \gws from \pbh inspirals with masses $\lesssim 0.1\msun$ could spend at least hours in the detector frequency range, which is problematic for matched-filtering  methods, since phase mismatch between templates accumulates with signal duration, thus requiring many more templates to cover the same parameter space \cite{Nitz:2021mzz}. Thus, we propose to search for such planetary-mass \pbhs with a more computationally efficient method than matched filtering: the \GFH \cite{Miller:2018rbg,Miller:2020vsl,Miller:2024jpo}.

Inspiraling planetary-mass \pbh binaries could lead to detectable \gwh signals if they formed within our galaxy, motivating the development of new methods to search for them \cite{Miller:2020vsl,Andrés-Carcasona:2023df,Alestas:2024ubs}. Furthermore, 
recent detections of star and quasar microlensing events from HSC \cite{Croon:2020ouk}, OGLE \cite{Niikura:2019kqi}, and EROS \cite{EROS-2:2006ryy} have suggested that \pbhs with masses between $10^{-6}M_\odot$ and $ 10^{-5}M_\odot$ could compose $\sim 2-10\%$ of \dm \cite{Hawkins:2020zie,bhatiani2019confirmation}. 
However, microlensing limits could be evaded if \pbhs form in clusters \cite{Garcia-Bellido:2017xvr,Calcino:2018mwh,Belotsky:2018wph,Carr:2019kxo,Carr:2019kxo,Trashorras:2020mwn,DeLuca:2020jug}, thus we must probe the same \pbh masses in different ways.  

In this work, we perform the first-ever search for \gws from planetary-mass \pbh binaries, and place the first \gwh constraints on the fraction of \dm that planetary-mass \pbhs could compose. Because our search is primarily sensitive to the chirp mass of the binary, we can constrain both equal-mass and asymmetric-mass ratio \pbh binaries, the latter of which may form more often. 

\section{Signal model}\label{sec:gwinsp}

Inspiraling compact objects will emit \gws that shrink their orbits over time, leading to their eventual merger \cite{maggiore2008gravitational}. When two objects are far from merger, we can approximate the \gwh emission as arising from two point masses orbiting around their center of mass. 
Equating orbital energy loss with \gwh power, the rate of change of the \gwh frequency over time (the spin-up) $\dot{f}_{\rm gw}$ is \cite{maggiore2008gravitational}:

\begin{equation}
    \dot{f}_{\rm gw}=\frac{96}{5}\pi^{8/3}\left(\frac{G\mathcal{M}}{c^3}\right)^{5/3} f_{\rm gw}^{11/3}\equiv k f_{\rm gw}^{11/3}.
    \label{eqn:fdot_chirp}
\end{equation}
$\mathcal{M}\equiv\frac{(m_1m_2)^{3/5}}{(m_1+m_2)^{1/5}}$ is the chirp mass of the system, $f_{\rm gw}$ is the \gwh frequency, $c$ is the speed of light, and $G$ is Newton's gravitational constant.  
Integrating \cref{eqn:fdot_chirp}, we obtain the frequency evolution $f_{\rm gw}(t)$:
\begin{equation}
f_{\rm gw}(t)=f_0\left[1-\frac{8}{3}kf_0^{8/3}(t-t_0)\right]^{-\frac{3}{8}}~,
\label{eqn:powlaws}
\end{equation}
where $t_0$ is a reference time for the \gwh frequency $f_0$ and $t$ is the time at $f_{\rm gw}$. 
The amplitude $h_0(t)$ from a source a distance $d$ away  evolves as \cite{maggiore2008gravitational}:

\begin{equation}
h_0(t)=\frac{4}{d}\left(\frac{G \mathcal{M}}{c^2}\right)^{5/3}\left(\frac{\pi f_{\rm gw}(t)}{c}\right)^{2/3}.
\label{eqn:h0}
\end{equation}

While \cref{eqn:fdot_chirp} represents the zeroth-order Post-Newtonian (PN) term \cite{maggiore2008gravitational}, because we observe the inspiral far from merger, our results are valid up to 3.5\pn for equal-mass and asymmetric mass ratio systems with $q=m_2/m_1\approx \eta \in [10^{-7},10^{-4}]$ in this parameter space \cite{Alestas:2024ubs} assuming $m_1\sim \mathcal{O}(\msun)$. 

\section{Search}\label{sec:search}

\subsection{Method}

{We begin with six months of cleaned, calibrated LIGO O3a strain data $h(t)$ \cite{Sun:2020wke,Virgo:2018gxa,Vajente:2019ycy, LIGOScientific:2020ibl,LIGO:2021ppb} that we divide into chunks of length $\TFFT$, which we fast Fourier transform (FFT). In each FFT, we estimate the level of the background noise power spectral density (PSD), and divide the square modulus of the FFT by the PSD to obtain the ``equalized power'' \cite{Miller:2024jpo}, which has mean and standard deviation equal to one in Gaussian noise \cite{Astone:2014esa}.}

{We repeat this procedure for each FFT to build a time-frequency peakmap, in which any ``peak'' (a time-frequency point) whose power has exceeded a threshold $\theta_{\rm thr}=2.5$ is labeled with ``1''. This peakmap is the input to the \GFH, which maps peaks in the time-frequency plane to lines in the frequency-chirp mass plane.}

\begin{figure*}[ht!]
    \centering
    \includegraphics[width =\linewidth]{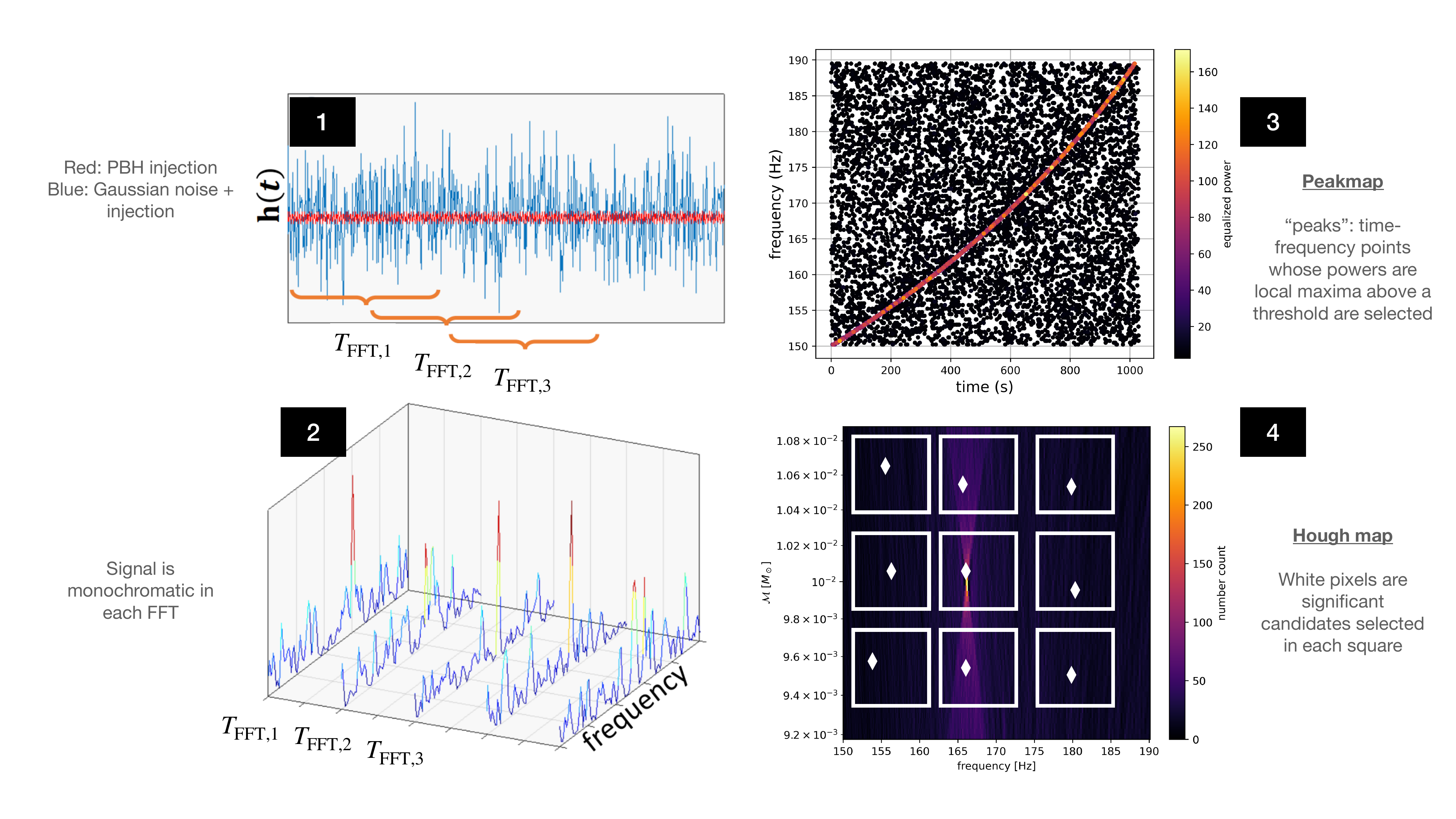}
    \caption{{Summary of our search for inspiraling \pbh binaries. Step 1: perform 50\% interlaced FFTs of length $\TFFT$, chosen such that the signal power is confined to one frequency bin in each FFT: $\dot{f}_{\rm gw}\TFFT\leq \frac{1}{\TFFT}$. Step 2: estimate the noise PSD and calculate the equalized power $\frac{|FFT|^2}{PSD}$ \cite{lorenzoposter}. Step 3: select local maxima above a threshold $\theta_{\rm thr}=2.5$ to build the time-frequency peakmap. Step 4: apply the \GFH to find the most likely $f_0$ and $\mathcal{M}$ of a \pbh binary, and select significant candidates in different ``squares'' of the Hough map.}}
    \label{fig:method}
\end{figure*}

{To apply the \GFH, we must linearize the peakmap by setting $z=f_{\rm gw}^{-8/3}$ in \cref{eqn:powlaws}, resulting in: $z=z_0-\frac{8}{3}k(t-t_0)$, where $z_0=f_0^{-8/3}$ \cite{Miller:2018rbg}. Our method sums the ones in the $t-z$ plane along different linear tracks, not the power, to avoid being blinded by noise disturbances \cite{Astone:2005fj,Astone:2014esa}, and results in a 2-dimensional histogram in the $z_0-k$ ($f_0-\Mc$) plane, called the ``Hough map''.} {We show a schematic of the search in \cref{fig:method}.}%

\begin{figure*}[ht!]
     \begin{center}
        \subfigure[ ]{%
            \label{fig:f0_Mc_dur_parmspace}

        \includegraphics[width=0.49\textwidth]{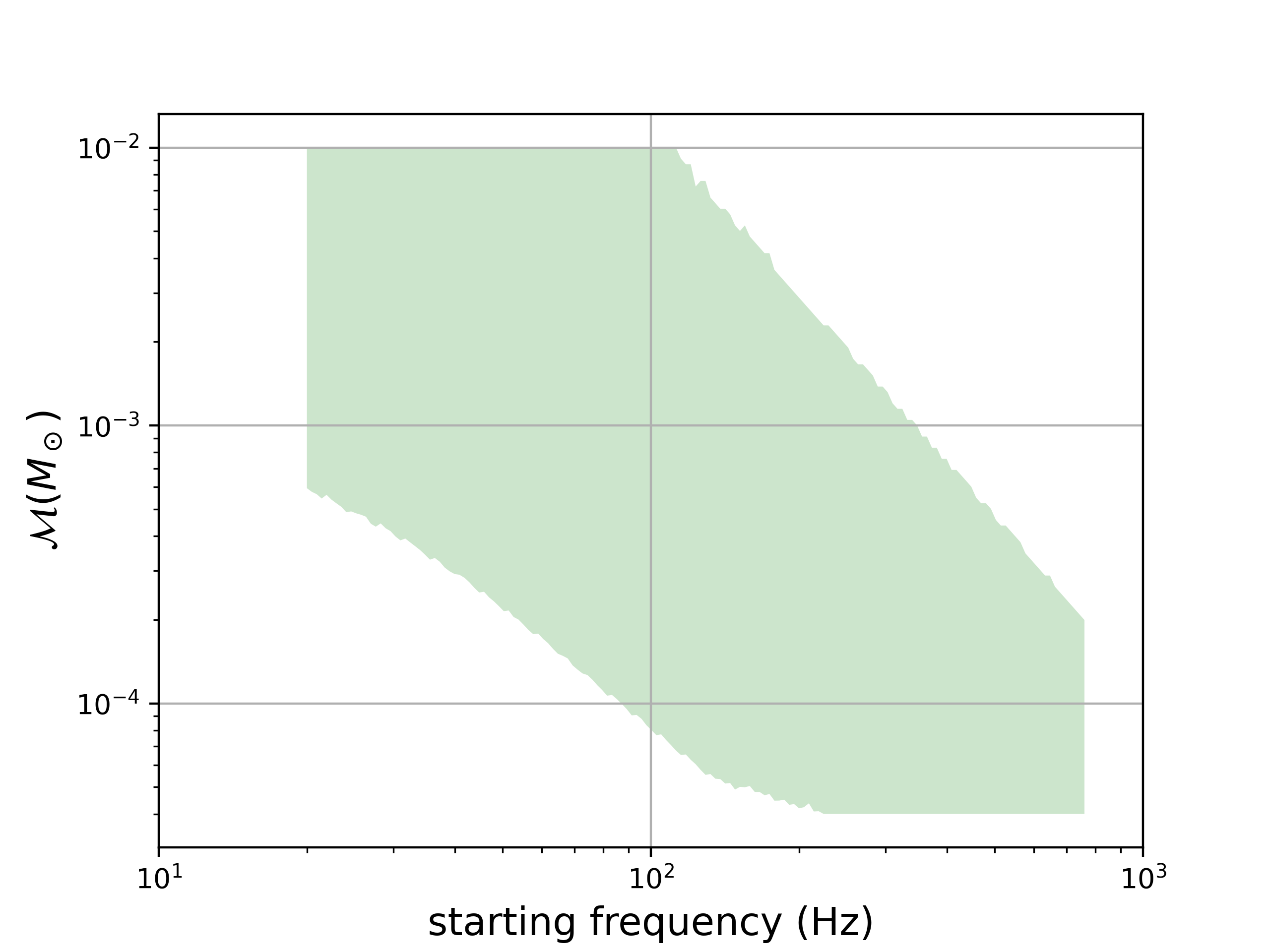}
        }%
        \subfigure[]{%
           \label{fig:tfft_fmin_fmax_parmspace}
           \includegraphics[width=0.49\textwidth]{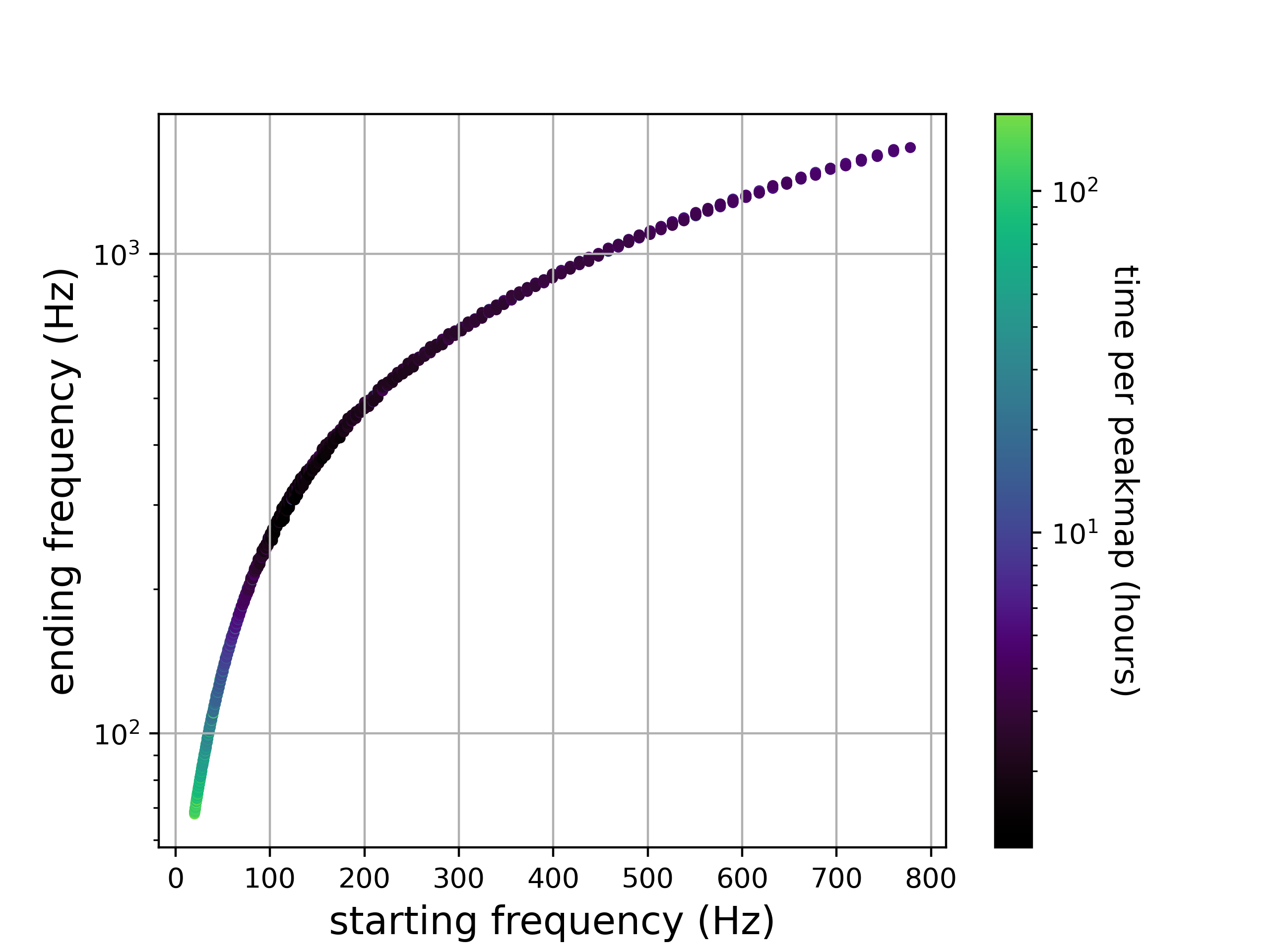}
        }\\ 
    \end{center}
    \caption[]{Left: the chirp mass range probed in our search as a function of $f_0$. The shape results because we analyze only systems that could be detected at least $0.1$ kpc away at 95\% confidence. Right: $\Tobs$ as a function of frequency range analyzed in this search. $\TFFT\in [2,29]$ s. }%
     \label{fig:parmspace}
\end{figure*}

We plot in \cref{fig:f0_Mc_dur_parmspace} the searched parameter space, which extends a few orders of magnitude lower in chirp mass than that in previous \ssm searches \cite{Nitz:2021vqh,LIGOScientific:2021job,LIGOScientific:2022hai}. In \cref{fig:tfft_fmin_fmax_parmspace}, we show $\Tobs$, the duration of each peakmap, as a function of starting and ending frequency of each analyzed band. We choose the frequency band, $\TFFT$ and $\Tobs$ {to maximize sensitivity to particular chirp-mass systems} \cite{Miller:2024jpo}, and thus run the search in $N_{\rm config}=129$ configurations \cite{miller_2024_10724845}. 
Each configuration corresponds to creating $N_{\rm PM}=T_{\rm obs}/\Tobs$ peakmaps and running the \GFH on each one. 


\subsection{Candidates and follow-up}

We analyze data from LLO and LHO separately, and {select strong ``candidates'', particular $z_0$ and $k$ values, whose number counts in the Hough map are high with respect to those in nearby $z_0-k$ pixels}, and find $\sim10^7$ coincident candidates. A ``coincidence'' occurs if the Pythagorean distance between the returned $z_0$ and $k$ of candidates from each detector are within 3 bins \cite{Miller:2018rbg} at the same $t_0$ \cite{Miller:2024jpo}. {For each candidate, we calculate the critical ratio $CR$, our detection statistic: $CR=\frac{m-\mu}{\sigma}$, where $m$ is the number count of each candidate in the Hough map, and $\mu$ and $\sigma$ are the mean and standard deviation, respectively, of the number counts in different squares of the Hough map \cite{Miller:2024jpo}.}

We define a configuration-dependent threshold $CR$ \cite{Miller:2024jpo} above which we decide candidates are ``significant'', and remove any candidates within a frequency bin of a known noise line \cite{Sun:2020wke}.
 {We determine if the remaining 334 candidates are real by returning to $h(t)$ and correcting the data for the phase evolution of these candidates, a process called ``heterodyning'' \cite{Piccinni:2018akm,Miller:2024jpo}. If the candidate parameters match the signal parameters in the data, heterodyning results in a monochromatic signal, allowing us to take a longer $\TFFT$, which leads to a larger $CR$. No candidate survived heterodyning.}

\section{Upper limits}

\subsection{Constraints on rate density}
We compute upper limits using a hybrid theoretical/data-driven approach that has been verified through injections \cite{Miller:2024jpo}.
The sensitivity of the \GFH search towards inspiraling binary systems has been computed in \cite{Miller:2020vsl} in terms of the maximum distance reach $d_{\rm max}$ at confidence level $\Gamma=0.95$:

\begin{eqnarray}
d_{\rm max}&=&\left(\frac{G \mathcal{M}}{c^2}\right)^{5/3}\left(\frac{\pi}{c}\right)^{2/3} \frac{\TFFT}{\sqrt{\Tobs}}\left(\sum_x^N \frac{f^{4/3}_{\text{gw},x}}{S_n(f_{\text{gw},x})}\right)^{1/2} \nonumber \\ &\times& \left(\frac{p_0(1-p_0)}{Np^2_1}\right)^{-1/4}\sqrt{\frac{\theta_{\rm thr}}{\left(CR-\sqrt{2}\erfc^{-1}(2\Gamma)\right)}}.
\label{eqn:dmax}
\end{eqnarray} 
Here, $p_0=e^{-\theta_{\rm thr}}-$  $e^{-2\theta_{\rm thr}}$ $+\frac{1}{3}e^{-3\theta_{\rm thr}}$ is the probability of selecting a peak above $\theta_{\rm thr}$, $p_1$ = $e^{-\theta_{\rm thr}}-$  2$e^{-2\theta_{\rm thr}}$ $+e^{-3\theta_{\rm thr}}$, $N=\Tobs/\TFFT$, and $S_n$ is the averaged LHO/LLO noise PSD in O3a, given in \cite{Miller:2024jpo}. $x$ indicates the sum over the theoretical frequency track for a system with chirp mass $\mathcal{M}$ and starting frequency $f_{\mathrm{gw},x=0}$.


{Even though each configuration was constructed to be sensitive to a particular $\mathcal{M}$ and $f_0$ \cite{Miller:2024jpo}, we can actually probe a wide range of chirp masses and starting frequencies in each Hough map. 
Within a given Hough map $j$ ($j\in[1,N_{\rm PM}]$) for a particular configuration $i$ ($i\in[1,N_{\rm config}]$), all coincident candidates selected with a chirp mass $\mathcal{M}_k$ and starting frequency $f_l$ are assigned a critical ratio ($CR_{i,j,k,l}$). Using the Feldman-Cousins approach to set upper limits, which ensures perfect coverage at a given confidence level, $CR_{i,j,k,l}$ is mapped to an ``inferred'' positive-definite $CR$ based on the upper value of Tab. 10 in \cite{Feldman:1997qc} at 95\% confidence. This approach produces consistent limits compared to those obtained by injecting simulated signals in real data \cite{Miller:2018rbg,Miller:2020kmv,KAGRA:2022dwb,Miller:2024jpo}. Because $CR_{i,j,k,l}$ is found for LHO and LLO separately, we conservatively use the maximum of the two in Eq. (\ref{eqn:dmax}) after applying the Feldman-Cousins approach.}

{Thus, we compute a distance reach $d_{i,j,k,l}$  for all $10^7$ candidates returned from the first step of the search. We then select $\sim 100$ chirp masses $\mathcal{M}_s$ that cover the searched parameter space at which we set upper limits, and, at each $\mathcal{M}_s$, in each configuration, take the median of the distance reaches over all frequencies and Hough maps (i.e. over $T_{\rm obs}$): $d_{i,s}=\text{median}_{j,l}(d_{i,k=s,j,l})$. Essentially, our procedure corresponds to the median sensitivity of the search over time.  }


We then have one distance reach per configuration per $\mathcal{M}_s$. Some configurations will be more sensitive than others to certain chirp masses, though they search overlapping chirp-mass ranges; therefore, we take the maximum distance for each chirp mass as the upper limit, i.e. $d_s = \max_i(d_{i,s})=d_{\rm max,95\%}(\mathcal{M}_s)$. 

We then compute the spacetime volume $\avgVT$ of nearby \pbh  binaries and their rate density $\mathcal{R}_{\rm 95\%}$ \cite{miller:2021knj,Alestas:2024ubs,Biswas:2007ni}

\begin{eqnarray}
    \avgVT &=& \frac{4}{3}\pi d_{\rm max,95\%}^3 T_{\rm obs}, \\
     \mathcal{R}_{\rm 95\%}&=&\frac{3.0}{\avgVT}.
\end{eqnarray}
where $T_{\rm obs}=$ \DaysDataAnalyzed (adjusted for the detector duty cycle of 70\%). 

\begin{figure*}[ht!]
    \centering
    \includegraphics[width =0.8\linewidth]{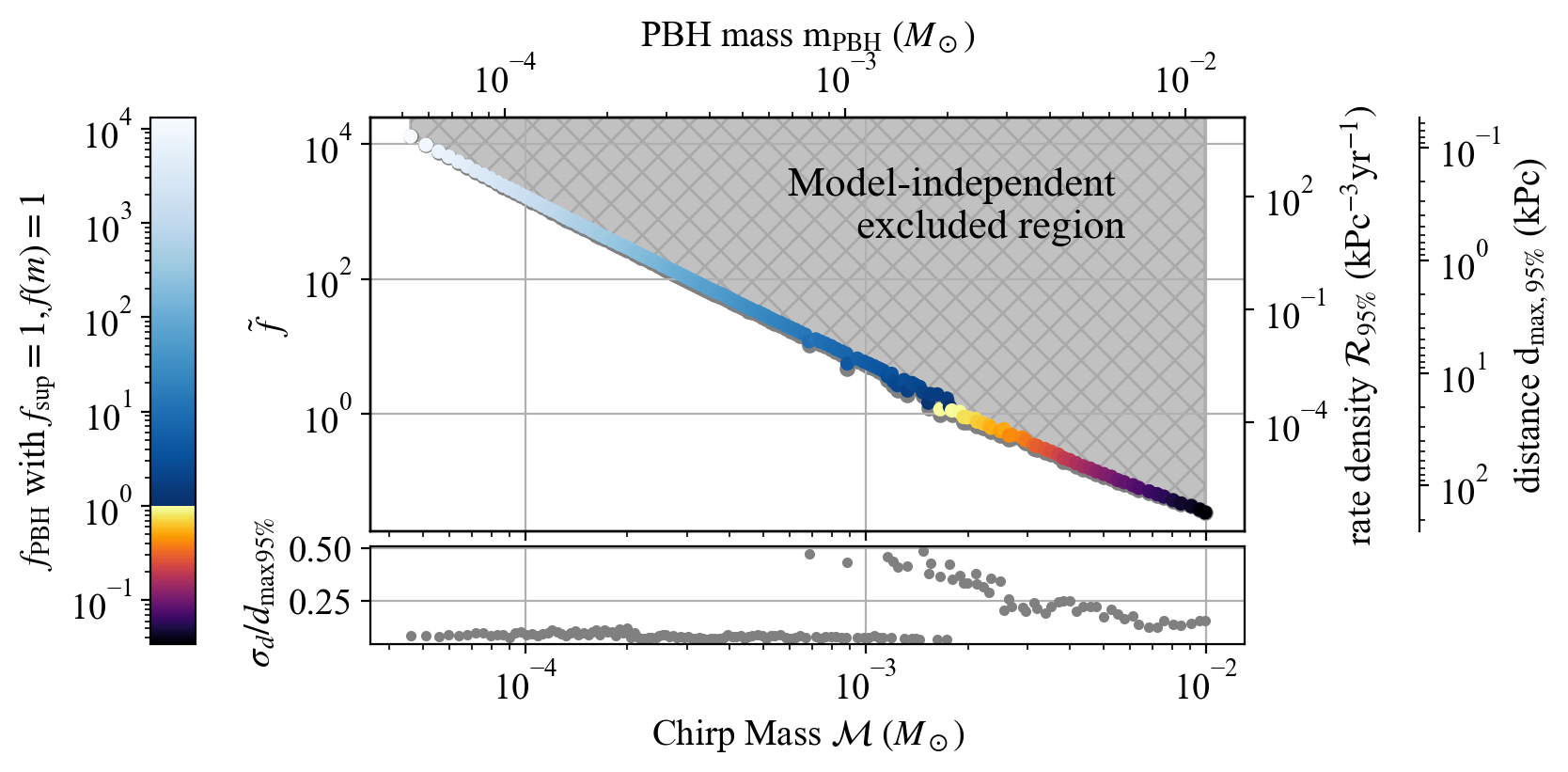}
    \caption{For equal-mass binaries, upper limits on $\tilde{f}$ (left y-axis), the merger rate density (right y-axis) and maximum distance reach (rightmost y-axis) as a function of chirp mass (lower x-axis) and PBH mass (upper x-axis). Gray-hatched regions denote model-independent constraints on $\tilde{f}$, distances and rate densities that are excluded by this analysis of O3a LIGO data. Model-dependent limits on $f_\mathrm{PBH}$ are shown on the color axis, assuming no rate suppression and a monochromatic mass function. We also include the fractional error on distance reached in the bottom sub-plot.  }
    \label{fig:equalmassnancy}
\end{figure*}

\begin{figure*}[ht!]
    \centering
    \includegraphics[width =0.8\linewidth]{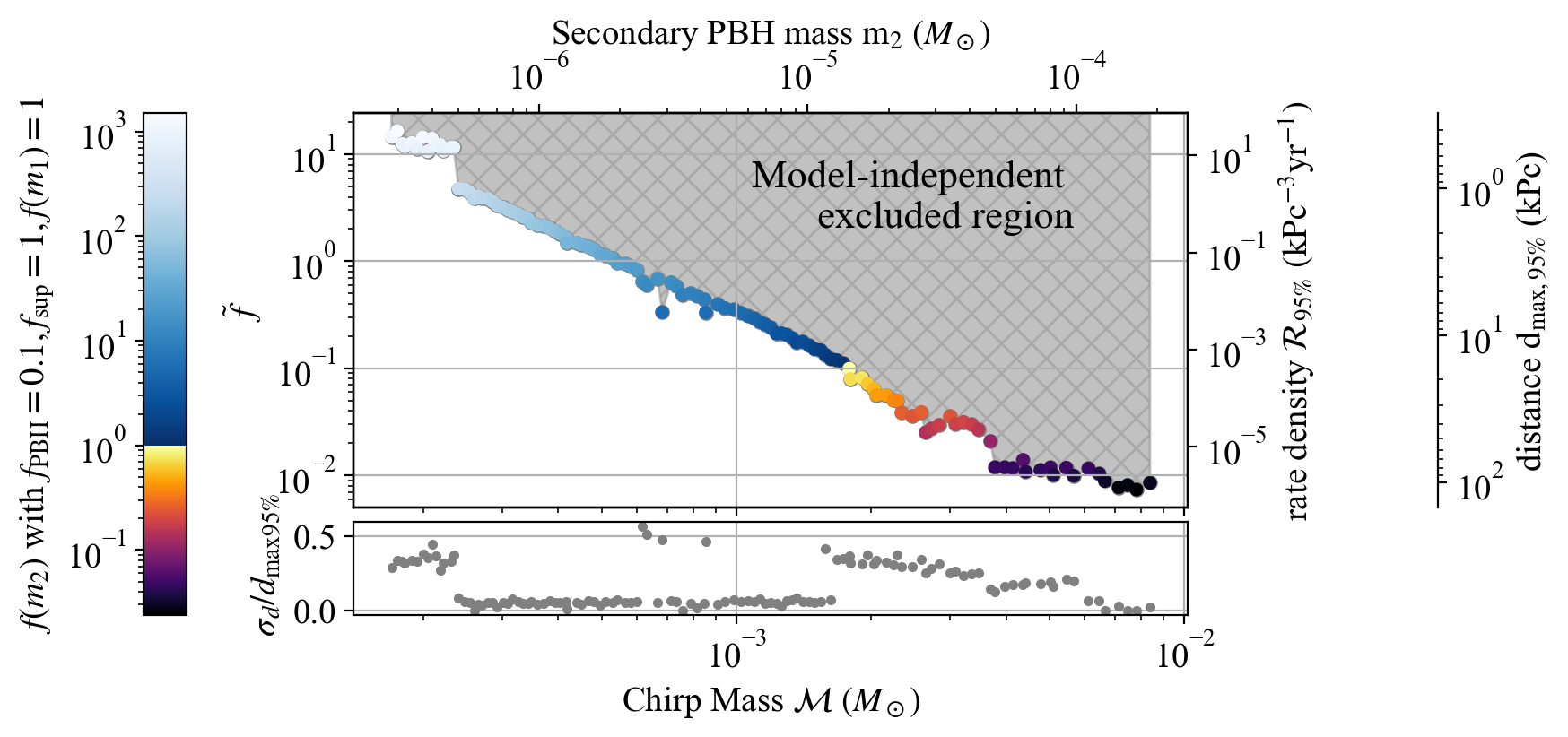}
    \caption{Same as \cref{fig:equalmassnancy}, but for asymmetric mass-ratio binaries. Model-dependent limits on $f(m_2)$ are shown on the color axis, assuming no rate suppression, $f(m_1)\sim1$, and $\fpbh=0.1$. }
    \label{fig:asymetricmassnancy}
\end{figure*}

The upper limits on $d_{\rm max,95\%}$ and $\mathcal{R}_{\rm 95\%}$ are shown in \cref{fig:equalmassnancy} and \cref{fig:asymetricmassnancy} for the equal-mass and asymmetric mass-ratio cases, respectively. Note that these figures look different because we only place limits when we can ensure that the 0\pn waveform does not differ by more than one frequency bin from the 3.5\pn waveform during $\Tobs$ \cite{Miller:2024jpo}. This makes asymmetric mass ratio systems harder to constrain than equal-mass ones, since the mass ratio enters at 1\pn \cite{Alestas:2024ubs}.

\subsection{Constraints on primordial black holes}

To specialize our results to \pbhs, we use the formulas from~\cite{raidal2019formation,Hutsi:2020sol} for the cosmological merger rates that assume a purely Poissonian \pbh spatial separation at formation, given by:
\bea
        \mathcal{R}^{\rm cos}_{\rm prim}  &\approx &  {1.6 \times 10^{-12}}{\text{ \invkpccubedyr}} f_{\rm sup} f_{\rm PBH}^{{53/37}} f(m_1) f(m_2) \nonumber \\  &\times & \left(\frac{m_1 + m_2}{M_\odot}\right)^{-32/37}\left[\frac{m_1 m_2}{(m_1+m_2)^2}\right]^{-34/37},  \label{eq:cosmomerg}
\eea
which correspond to the rate per unit of logarithmic mass of the two binary black hole components $m_1$ and $m_2$. 
$f(m)$ is the mass distribution function of \pbhs normalized to one, and $f_{\rm sup}$ accounts for rate suppression due to the gravitational influence of early forming \pbh clusters, nearby \pbhs and matter inhomogeneities~\cite{raidal2019formation}. 
As in \cite{Miller:2020vsl}, we calculate the expected merging rates locally by assuming a constant local \dm density of $\sim 10^{16} M_\odot {\rm Mpc}^{-3}$ \cite{Weber:2009pt} consistent with the galactic \dm profile. In this way, we obtain $\mathcal{R} = 3.3 \times 10^5 \mathcal{R}_{\rm prim}^{\rm cos}$~\cite{Miller:2020kmv}.

Values for $\fsup$ vary depending on how wide the mass function is, how high the mass ratio and eccentricity of the binary are, and whether external tidal fields affect binary evolution \cite{Eroshenko:2016hmn,Cholis:2016kqi,Hutsi:2020sol,Clesse:2020ghq,raidal2019formation}. We therefore provide limits on an effective parameter $\tilde{f}$, defined to be model-agnostic:

\begin{equation}
\tilde f^{53/37} \equiv f_{\rm sup} f(m_1) f(m_2) f_{\rm PBH}^{53/37}.\label{ftilde}
\end{equation}

First, we constrain equal-mass \pbh merger rate densities and denote this mass $m_{\rm PBH}$, for which 
\begin{eqnarray}
\mathcal{R} =& 1.04 \times 10^{-6}\, \mathrm{kpc}^{-3} \mathrm{yr}^{-1} \left(\frac{m_\mathrm{PBH}}{M_\odot}\right)^{-32/37} \tilde{f}_{\rm equal}^{53/37}.
\label{eqn:rate}
\end{eqnarray}
Upper limits on $\tilde{f}$ are shown in \cref{fig:equalmassnancy}, which are the first to probe $\tilde{f}<1$. We also provide constraints on $f_\mathrm{PBH}$ assuming a monochromatic mass function and no rate suppression, which is below 1 for $\mpbh\gtrsim 2\times 10^{-3}M_\odot$.

We can also place constraints on systems with highly asymmetric mass ratios, assuming negligible eccentricity, for which formation rates are larger by several orders of magnitude. Since \pbhs are well-motivated by observations of merging black holes in the stellar-mass range, we consider the merging rate densities of systems with $m_1 = 2.5\, M_\odot$~\cite{Byrnes:2018clq,Carr:2019kxo,Clesse:2020ghq,Clesse:2020ghq}, and $m_2\ll m_1$:
\begin{eqnarray}
\mathcal{R} &=& 5.28 \times 10^{-7}\, \mathrm{kpc}^{-3} \mathrm{yr}^{-1} \left(\frac{m_1}{M_\odot}\right)^{-32/37} \nonumber \\ &\times& \left(\frac{m_2}{m_1}\right)^{-34/37} \tilde{f}_{\rm asymm}^{53/37},
\label{eqn:rate_asymm}
\end{eqnarray}
and constrain \(\tilde f_{\rm asymm}\) as a function of \(m_2\) in \cref{fig:asymetricmassnancy}.

To constrain $f(m_2)$, we assume that the mass function is dominated by $m_1$, i.e. $f(m_1) \approx 1$, which would be expected for broad mass functions affected by the QCD transition, implying $\tilde f = f_{\rm PBH} [f(m_2)]^{37/53} f_{\rm sup}^{37/53}$.  If we assume a model in which $\fpbh=0.1$ and $\fsup=1$, we can translate these limits to constrain $f(m_2)$, which are also shown in \cref{fig:asymetricmassnancy}. We see exclusions ($f(m_2)<1$) when $m_2\ge10^{-5}M_\odot$.

\section{Conclusions}

We presented results from the first-ever search for \gws from inspiraling planetary-mass \pbh binaries. Depending on the chirp mass, we could detect binary \pbhs that formed between [0.1,100] kpc away. Furthermore, we show that $\fpbh<0.1$ for equal-mass \pbh binaries with $\mpbh\sim[5\times 10^{-3},2\times 10^{-2}]\msun$, assuming no rate suppression and monochromatic mass functions. For asymmetric mass ratio systems, we constrain $f(m_2)<0.1$ if $m_1=2.5\msun$ and $m_2\gtrsim 1.5\times 10^{-5}\msun$, assuming $\fpbh=0.1$. These limits are conservative with respect to those obtained with injections \cite{Miller:2024jpo}. Furthermore, a major benefit of our formulation of upper limits is that they can be applied to \emph{any} exotic compact object formation models, e.g. strange quark star inspirals \cite{Geng:2015uja}, or a \pbh inspiral inside a compact object \cite{Zou:2022wtp}.


However, our results neglect eccentricity of the binary, which could induce changes in the waveform, meaning that we would need to be modify our method, or instead employ a model-agnostic one \cite{Alestas:2024ubs}. Furthermore, our limits on $\fpbh$ and $f(m_2)$ should be reevaluated if non-Poissonian clustering, e.g. from primordial non-Gaussianity in the initial curvature distribution, occurs. Our work motivates future searches of planetary- and asteroid-mass \pbh binaries, and our results will continue to improve when new data becomes available.

\section*{Acknowledgments}
This material is based upon work supported by NSF's LIGO Laboratory which is a major facility fully funded by the National Science Foundation

We would like to thank the Rome Virgo group for the tools necessary to perform these studies, such as the development of the original \fh transform and the development of the short FFT databases. Additionally we would like to thank Luca Rei for managing data transfers.

This research has made use of data, software and/or web tools obtained from the Gravitational Wave Open Science Center (https://www.gw-openscience.org/ ), a service of LIGO Laboratory, the LIGO Scientific Collaboration and the Virgo Collaboration. LIGO Laboratory and Advanced LIGO are funded by the United States National Science Foundation (NSF) as well as the Science and Technology Facilities Council (STFC) of the United Kingdom, the Max-Planck-Society (MPS), and the State of Niedersachsen/Germany for support of the construction of Advanced LIGO and construction and operation of the GEO600 detector. Additional support for Advanced LIGO was provided by the Australian Research Council. Virgo is funded, through the European Gravitational Observatory (EGO), by the French Centre National de Recherche Scientifique (CNRS), the Italian Istituto Nazionale della Fisica Nucleare (INFN) and the Dutch Nikhef, with contributions by institutions from Belgium, Germany, Greece, Hungary, Ireland, Japan, Monaco, Poland, Portugal, Spain.

We also wish to acknowledge the support of the INFN-CNAF computing center for its help with the storage and transfer of the data used in this paper.

We would like to thank all of the essential workers who put their health at risk during the COVID-19 pandemic, without whom we would not have been able to complete this work.

F.D.L. is supported by a FRIA (Fonds pour la formation à la Recherche dans l'Industrie et dans l'Agriculture) Grant of the Belgian Fund for Research, F.R.S.-FNRS (Fonds de la Recherche Scientifique-FNRS).

This work is partially supported by ICSC – Centro Nazionale di Ricerca in High Performance Computing, Big Data and Quantum Computing, funded by European Union – NextGenerationEU.

\bibliographystyle{apsrev4-1}
\bibliography{biblio,biblio_pbh_method}

\end{document}